\RequirePackage[2020/08/10]{latexrelease}
\documentclass[aps,pra,reprint,superscriptaddress,longbibliography]{revtex4-2}

\usepackage{physics}
\usepackage{amsmath}
\usepackage{amssymb}
\usepackage{graphicx}
\usepackage{float}
\usepackage{mathrsfs}
\usepackage{mathtools}
\usepackage{extarrows}
\usepackage[margin=0.7in]{geometry}
\graphicspath{{PNG/}}
\usepackage{xcolor}
\newcommand{\RNum}[1]{\uppercase\expandafter{\romannumeral #1\relax}}

\begin{document}
    
	\title{Quench dynamics of edge states in a finite extended Su-Schrieffer-Heeger system}% Force line breaks with \\
	
	\author{A.~ Ghosh}
		
	\affiliation{Indian Institute of Technology Kharagpur, Kharagpur 721302, West Bengal, India}
	
	\affiliation{School of Physics, University of Melbourne, Melbourne, 3010, Australia}
	
	\author{A.~M.~Martin}
	
	\affiliation{School of Physics, University of Melbourne, Melbourne 3010, Australia}

	\author{S.~Majumder}
	
	\affiliation{Indian Institute of Technology Kharagpur, Kharagpur 721302, West Bengal, India}
	
	\date{\today}

	\begin{abstract}
		
	We examine the quench dynamics of an extended Su-Schrieffer-Heeger(SSH) model involving long-range hopping that can hold multiple topological phases. Using winding number diagrams to characterize the system's topological phases geometrically, it is shown that there can be multiple winding number transition paths for a quench between two topological phases. The dependence of the quench dynamics is studied in terms of the survival probability of the fermionic edge modes and post-quench transport. For two quench paths between two topological regimes with the same initial and final topological phase, the survival probability of edge states is shown to be strongly dependent on the winding number transition path. This dependence is explained using energy band diagrams corresponding to the paths. Following this, the effect of the winding number transition path on transport is investigated. We find that the velocities of maximum transport channels varied along the winding number transition path. This variation depends on the path we choose, i.e., it increases or decreases depending upon the path. An analysis of the coefficient maps, energy spectrum, and spatial structure of the edge states of the final quench Hamiltonian provides an understanding of the path-dependent velocity variation phenomenon. 
		
	\end{abstract}
	
	\maketitle

	\section{Introduction \label{sec:1}}
Quench dynamics is an area of considerable interest in topological systems research \cite{sacramento2016edge, wang2018detecting, budich2016dynamical, yang2018dynamical, zhang2018dynamical, rajak2014survival, rajak2017survival, sacramento2014fate}. Topics such as identifying topological features of dynamical quantum phase transitions \cite{budich2016dynamical, yang2018dynamical}, understanding how equilibrium topological quantum phases can be characterized in non-equilibrium scenarios \cite{zhang2018dynamical} and investigating how information spreads in many body systems after a quench [2,5] have been studied. For example, studies have shown that quasi-particle correlations spread when a system is quenched from deep inside a Mott insulating phase to the boundary of the superfluid phase \cite{cheneau2012light, bernier2018light}, and research on XXZ Heisenberg chains has provided insights into how correlations spread in interacting many-particle systems \cite{bonnes2014light}.

One of the main thrusts in recent years has been to study the post-quench fate of edge states in various topological settings, revealing some intriguing non-equilibrium physics. For instance, one-dimensional topological systems may contain isolated Majorana and Fermionic modes, which exhibit fascinating behavior of the survival probabilities \cite{sacramento2016edge}. It has also been found that the post-quench dynamical behavior of Majorana modes can also characterize the topological phase present in the system \cite{wang2018detecting}. When quenched to quantum critical points, the survival probability of p-wave edge modes oscillates depending on the system configuration \cite{rajak2014survival, rajak2017survival, sacramento2014fate}. Quench dynamical studies are performed on the Kitaev one-dimensional superconductor \cite{kitaev2001unpaired} with p-wave pairing and one-dimensional dimerized Kitaev superconductors \cite{wakatsuki2014fermion}. The dimerized Kitaev model can host several topological phases where the fermionic modes appear in the Su-Schrieffer-Heeger (SSH)-type Hamiltonian, which arises as a special case in the dimerized Kitaev model. The SSH system being the simplest topological system \cite{asboth2016short, su1979solitons}, plays an important role in most of these quench studies, and there have also been several interesting works on its experimental realization \cite{kiczynski2022engineering, meier2016observation, liu2022ta, cooper2019topological}. Considering that SSH systems can host multiple higher topological states depending on the long-range interaction, there is a need for understanding the dynamics of quench between these states.

In this paper, we study the SSH Hamiltonian in the presence of long-range hopping. For such a system, we geometrically show that it can host multiple topological phases. Based on the details of long-range hopping, the corresponding topological state is characterized by a winding number. Specifically, as the range of hopping increases, the maximum winding number the system can hold increases accordingly. We then investigate the quench dynamics of the system in terms of the survival probabilities of Fermionic edge modes corresponding to the topological phase. We examine the survival probability of an edge mode, in a topological state, when the system is quenched to another topological state. Surprisingly,  if we start in a well-defined edge state and quench via two different paths to a final topological state, the survival probability can be different, even though the topology of the system pre- and post-quench is the same. 
Following that, we examine the spread of probability density over time on each site and observe the dynamical behavior of the quench, through "light-cone" diagrams. The slope of a channel in the light cones gives the velocity of that channel. Furthermore, we explain how the velocity of maximum transport varies along a quench path and how this variation is dependent on the path chosen.

The paper is organized as follows. In Section \RNum{2}, we introduce the extended SSH Hamiltonian with up to fourth nearest neighbor tunneling, enabling the possibility of having a maximum winding number of four. In this section, using a geometric approach to mark the topological state, we introduce the concept of the path of winding number transition, and we show that there can be multiple paths of winding number transition possible corresponding to the same set of initial and final quench topologies. Section \RNum{3} presents our study of the dependence of quench dynamics of the fermionic edge modes on different paths of winding number transition corresponding to the same set of initial and final quenched topological states, with the results being explained via energy band diagrams corresponding to the quench paths. Following that, we demonstrate the variation in transport light cones corresponding to the quenches along these paths. By using coefficient maps of basis states, the energy spectrum of the final Hamiltonian and spatial structuring of the edge states of the final Hamiltonian, we proceed to explain the dependence of velocity variation of the transport on the path. Finally, in Section \RNum{4} we review our findings and consider future avenues of possible research.

	\section{Extended SSH system \label{sec:2}}
%	\subsection{Extended SSH Hamiltonian}
%	\begin{figure}[H]
%		\centering
%		\includegraphics[width=0.7\linewidth]{"work 1/extended SSH cartoon"}
%		\caption{Schematic diagram of extended SSH topological wire with each box representing an unit cell with two sublattices A (hollow circle) and B(solid circle). }
%		\label{fig:extended-ssh-cartoon}
%	\end{figure}
	
In this work, we consider a dimensionless extended SSH Hamiltonian  i.e., a SSH system with long-range hopping present: 
\begin{eqnarray}
		\hat{H}&=& v \sum_{m=1}^{N}|m,A\rangle\langle m,B| + w\sum_{m=1}^{N-1}|m+1,A\rangle\langle m,B| \nonumber \\
		&+&\nu\sum_{m=1}^{N-2}|m+2,A\rangle\langle m,B|+\mu\sum_{m=1}^{N-3}|m+3,A\rangle\langle m,B|\nonumber \\
		&+&\gamma\sum_{m=1}^{N-4}|m+4,A\rangle\langle m,B| +{\rm h.c}, 
\end{eqnarray} 
where, $v$ is the dimensionless intra-cellular hopping term, and $w$, $\nu$, $\mu$ and $\gamma$ are the dimensionless inter-cellular first, second, third, and fourth nearest neighbor hopping terms. These hopping terms respect the chiral symmetry i.e., they represent hopping between two different sub-lattices ($A$ and $B$ in this case). This system can host several topological phases. We use winding number diagrams to keep track of the topological phases in a fashion similar to the works in \cite{zhang2015topological, yin2018geometrical}. To plot the winding number diagram for the system, we  express $\hat{H}$ in terms of bulk momentum basis:
\begin{eqnarray} 
 \hat{H} (k)=\langle k|\hat{H}|k\rangle = \sum_{\alpha,\beta \in \left\lbrace A,B \right\rbrace } \bra{k,\alpha} \hat{H} \ket{k,\beta} \ket{\alpha}\bra{\beta},
 \end{eqnarray}
 where the plane wave basis $|k\rangle$   is given by \cite{asboth2016short}
\begin{eqnarray}
|k\rangle=\frac{1}{\sqrt{N}}\sum_{m=1}^{N}e^{imk}|m\rangle		
\end{eqnarray} 
and 
 \begin{eqnarray}
\hat{H}(k)=\begin{bmatrix}	0 & a(k) \\
a^{\dagger}(k) & 0 \end{bmatrix},
\end{eqnarray} 
where $a(k)=v+we^{ik}+\nu e^{i2k} + \mu e^{i3k}+\gamma e^{i4k}$. The matrix $\hat{H}(k)$  can be expressed in terms of Pauli matrices as
\begin{eqnarray}
\hat{H}(k)=d_0\hat{\sigma}_0 + d_x\hat{\sigma}_x + d_y\hat{\sigma}_y + d_z\hat{\sigma}_z,
\end{eqnarray} 
where,  $\hat{\sigma}_0$, $\hat{\sigma}_x$,  $\hat{\sigma}_y$ and $\hat{\sigma}_z$  are the Pauli matrices, and
\begin{eqnarray}
d_x&=&v+w\cos(k)+\nu\cos(2k) + \mu\cos(3k) + \gamma \cos(4k),  \nonumber \\
\\
d_y&=&w\sin(k)+\nu\sin(2k) + \mu\sin(3k) + \gamma \sin(4k), \\
d_0&=&d_z=0.
\end{eqnarray} 
The winding number of the system is defined as the number of times the tip of the ${\bf d}$ vector encompasses the origin of ($d_x,d_y$)-space as $k$ varies from $0$ to $2 \pi$.

\begin{figure}
	\centering
	\includegraphics[width=0.9\linewidth]{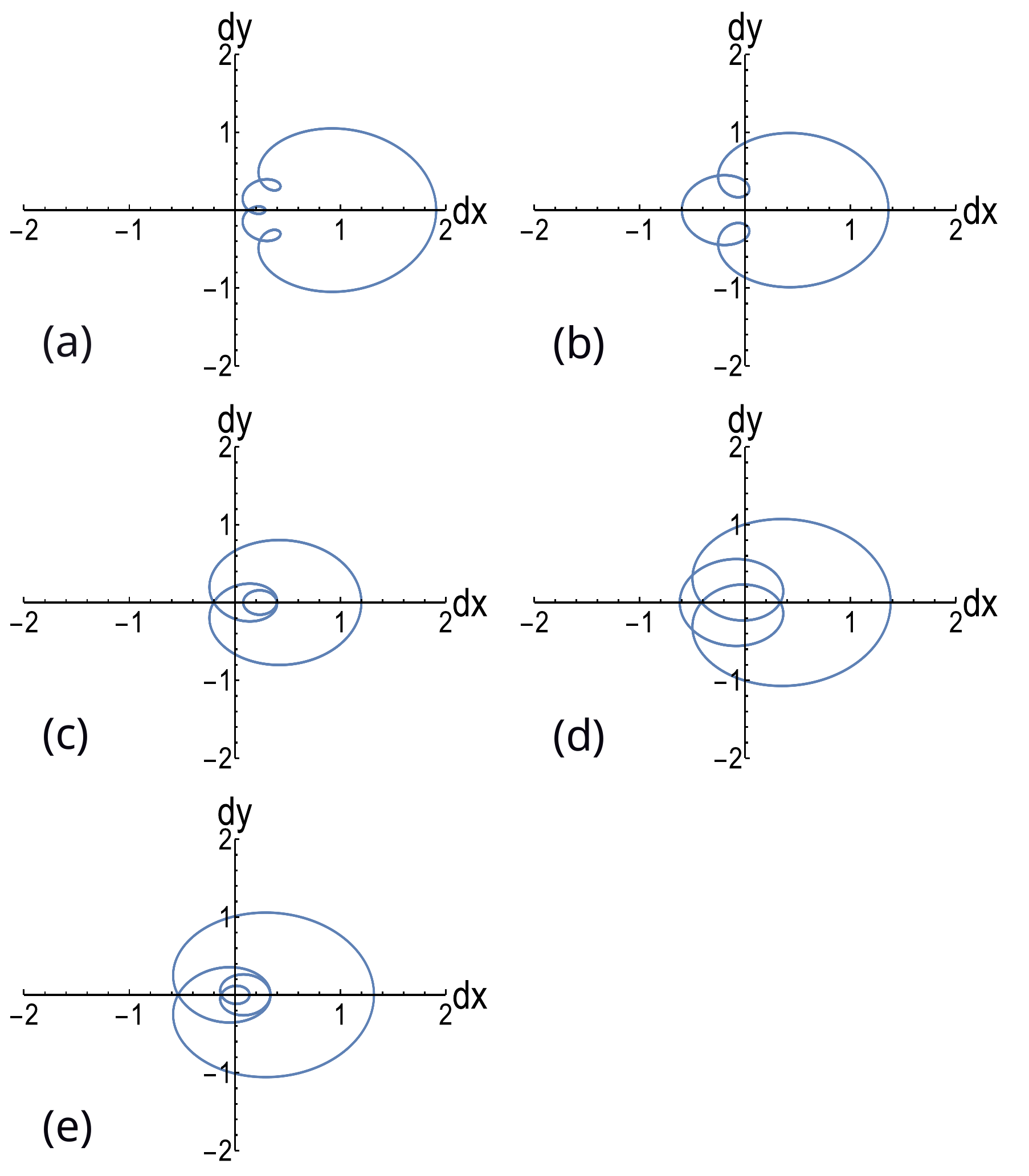}
	\caption{The winding number representation in $d_x - d_y$ plane. (a), (b), (c), (d) and (e) represent some possible topological states of the system with winding numbers 0, 1, 2, 3, and 4 respectively. The hopping parameters $ v=0.52 $, $ w= 0.54 $, $ \nu= 0.34 $, $ \mu= 0.27 $, and $ \gamma= 0.24 $ in (a); $ v= 0.1 $, $ w = 0.6 $, $ \nu = 0.28 $, $ \mu =0.38 $, and $ \gamma = 0 $ in (b); $ v = 0.28 $, $ w = 0.2 $, $ \nu = 0.36 $, $ \mu = 0.36 $, and $ \gamma = 0 $ in (c); $ v = 0.1 $, $ w = 0.4 $, $ \nu = 0.28 $, $ \mu = 0.6 $, and $ \gamma = 0 $ in (d); $ v = 0.1 $, $ w = 0.17 $, $ \nu = 0.25 $, $ \mu = 0.42 $, and $ \gamma = 0.38 $ in (e).}
	\label{fig:figure1}
		\vspace{-0.5cm}
\end{figure}

Figure \ref{fig:figure1} shows the winding number diagram corresponding to the topological phases possible in the system for different sets of parameters $v$, $w$, $\nu$, $\mu$, and $\gamma$. It is important to note that these diagrams are not unique for the winding numbers they are representing \cite{asboth2016short}. However, an important characteristic of the topological phases is that in a finite system, they host zero energy edge states or fermionic edge modes in this case. In an ideal SSH system without any domains, the number of edge states is two times the winding number, a topological invariant.

\subsection{The path of winding number transition}
In our study, the system is prepared in an initial state corresponding to an initial quench configuration with a particular topological phase characterized by the winding number, $C_i$. The system is then quenched to a final quench configuration, with a topological phase $C_f$. 
\begin{figure}
	\centering
	\includegraphics[width=0.9\linewidth]{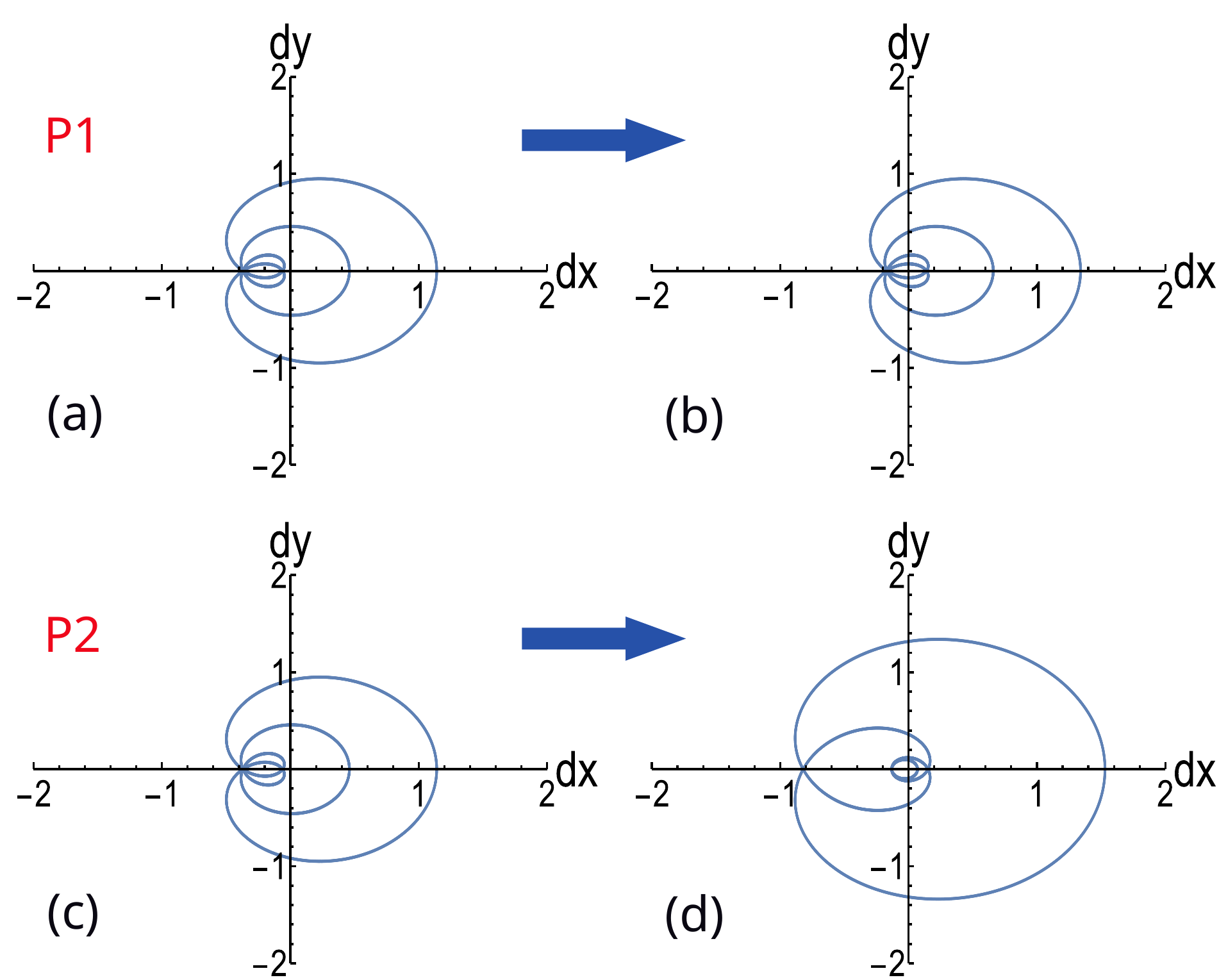}
	\caption{Two possible paths for a transition from winding number $2$ to winding number $4$. $P1$ corresponds to a change in parameter $v$ from $0$ to $0.2$, whilst keeping $w$, $\nu$, $\mu$, and $\gamma$ fixed at $0.17$, $0.43$, $0.17$, and $0.37$ respectively. $P2$ corresponds to a change in parameter $\mu$ from $0.17$ to $0.56$, keeping the other parameters $v$, $w$, $\nu$ and $\gamma$ fixed at $0$, $0.17$, $0.43$, and $0.37$, respectively.}
	\label{fig:figure2}
		\vspace{-0.5cm}
\end{figure}

\begin{figure*}
	\centering
	\includegraphics[width=0.9\linewidth]{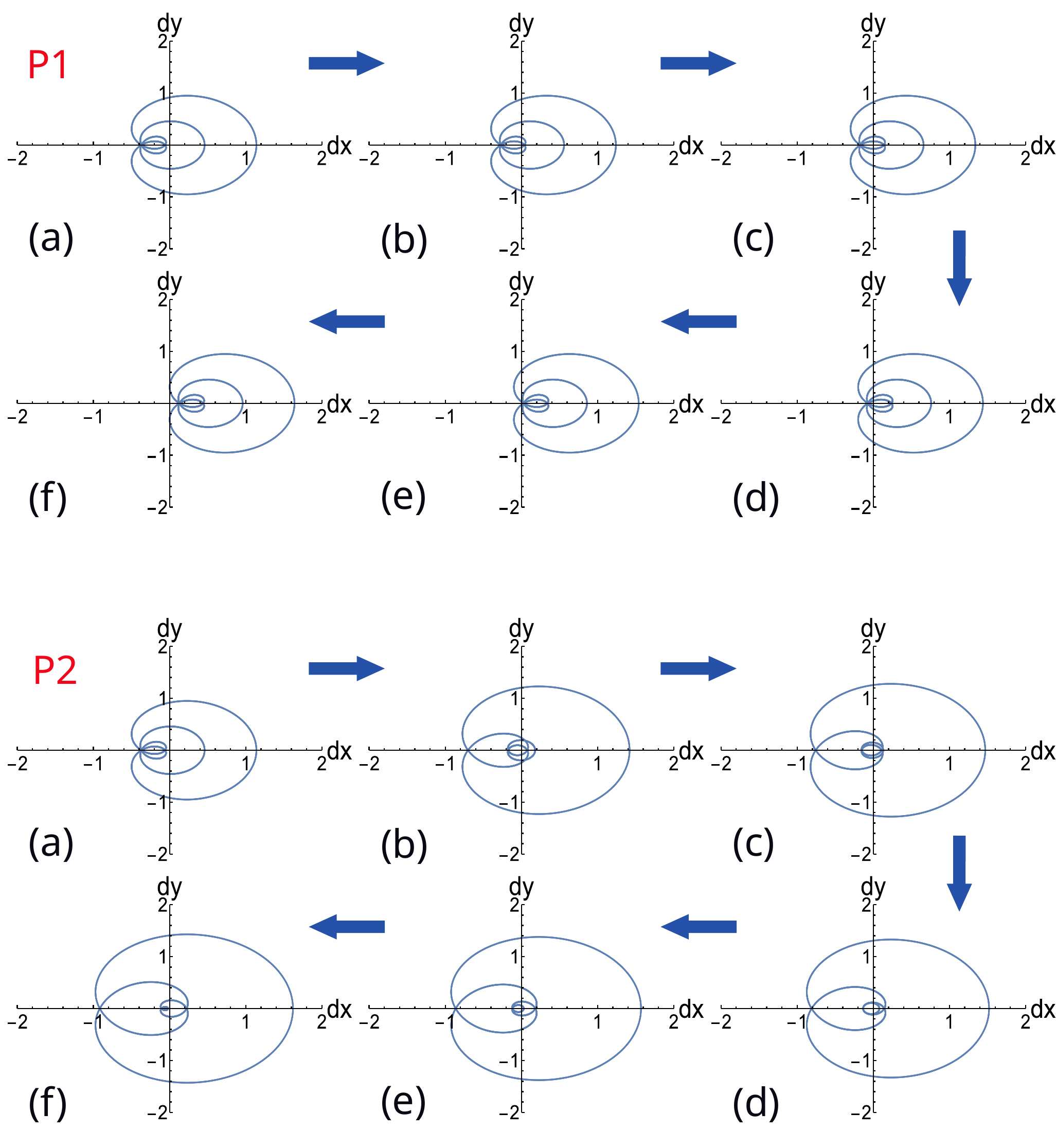}
	\caption{ The topological phases that the system takes along the two paths of winding number transition corresponding to the same initial winding number transition, from $2$ to $4$. In path $P1$, the parameter $v$ is gradually changed from $0$ to $0.5$, keeping other parameters $w$, $\nu$, $\mu$, and $\gamma$ fixed at $0.17$, $0.43$, $0.17$, and $0.37$, respectively.(a), (b), (c), (d), (e), and (f) of $P1$ corresponds to the value of $v$ as $0.0$, $0.1$, $0.2$, $0.3$, $0.4$, and $0.5$ respectively. In P1, the topological phase flow is 2-4-0. In $P2$ the parameter $\mu$ is increased from $0.17$ to $0.65$ keeping other parameters $v$, $w$, $ \nu$, and $\gamma$ at $0$, $0.17$, $0.43$, and $0.37$, respectively. For $P2$, (a), (b), (c), (d), (e), and (f) corresponds to the value of the parameter $\mu$ as $0.17$, $0.45$, $0.5$, $0.55$, $0.6$, and $0.65$, respectively. In P2, the topological phase flow is 2-4-3.}
	\label{fig:figure3}
	\vspace{-0.5cm}
\end{figure*}

Changing a parameter may change the winding number of the system, i.e., it may change from one winding number (say $C_i$) to another winding number (say $C_f$). It is also possible to bring about this same transition of winding number ($C_i$ to $C_f$) by changing some other parameter. These ways of manifesting the same winding number transition constitute the {\it path of winding number transition}. In Fig.~\ref{fig:figure2}, we provide an example of such multiple paths by demonstrating two possible paths for the transition from the topological phase with  $C_i=2$ to $C_f=4$. The first path ($P1$) corresponds to the change of the parameter $v$ from $0$ to $0.2$, keeping the other parameters $w$, $\nu$, $\mu$, and $\gamma$ fixed at $0.17$, $0.43$, $0.17$, and $0.37$, respectively. The second path ($P2$) corresponds to the change of the parameter $\mu$ from $0.17$ to $0.56$, keeping the other parameters $v$, $w$, $\nu$ and $\gamma$ fixed at $0$, $0.17$, $0.43$, and $0.37$, respectively. In each case, the topology of the initial and final states are the same, respectively. In essence, the question we wish to ask and answer in section III is the following. Suppose one starts in an edge state associated with a particular topology, and the system is quenched to another topological state. Does the winding number transition path impact the quench dynamics in terms of post-quench transport and survival probability of an edge state? Put another way, is the quench dynamics of an edge state different for the paths $P1$ and $P2$ shown in Fig.~\ref{fig:figure2}? If it is different, what governs the properties of the survival probability of the edge state and post-quench transport ("Light cones") through the system when quenching between two topological states?

Since we are investigating the dynamics of the quench along the paths corresponding to the various paths, it is necessary to investigate the topological phases that the system {\it passes} through along and past the quench. As an example, Fig. \ref{fig:figure3} denotes the topological phases that the system passes through and past for the two paths $P1$ and $P2$, shown in Fig.~\ref{fig:figure2}. For the first path, the topological phases along the path are $2$, $4$, but if the $v$ continues past $0.4$, the system ends up in the topological state $0$ ($2-4-0$). For the second path, it is $2$, $4$ and $3$ ($2-4-3$), where, as $\mu$ continues past $0.6$, the system ends up in the topological state $3$. We will see in Section III that one of the most critical aspects of this characterization of a path ($ 2-4 $) is where the topological phase for the quench is heading if we continue along the path of winding number transition: for $ P1 $ it is $ 0 $ and for $ P2 $ it is $ 3 $.%\begin{figure}
%	\centering
%	\includegraphics[width=0.7\linewidth]{"E:/My works/Figures/work 1/Figure 4"}
%	\caption{Showing the topological phases which the system takes along the two paths of w.n transition corresponding to the same initial w.n transition $2-3$. In the path $P1$, the parameter $w$ is gradually changed from $0.14$ to $0.6$ keeping the other parameters $v$, $\nu$, $ \mu$ and $\gamma$ fixed at $0.13$, $0.45$, $0.4$ and $0$ respectively. In the path $P2$, the parameter $\mu$ is gradually changed from $0.4$ to $0.9$ keeping other parameters $w$, $v$, $\nu$ and $\gamma$ fixed at $0.13$, $0.14$, $0.45$ and $0$ respectively}
%	\label{fig:figure-4}
%\end{figure}

%\begin{figure}
%		\centering
%		\includegraphics[width=6cm, height=8.5cm,angle=0] {Atomlaser_regions19.pdf}
%		\caption{Schematic of the atom laser operation in three different regions: (I), (II) and (III) indicate the WKB inside the BEC, Kirchhoff and paraxial regimes respectively. The solid blue sphere shows the BEC while the solid dark blue arrow along the $z$-axis depicts the atom laser beam. The harmonic focusing potential has been illustrated from the top view ($x-z$ plane) by the two shaded oval areas at the bottom of the figure. The intensity of the potential reaches its maximum value at the black spots whereas the minimum intensity occurs between the two peaks.}
%	\end{figure}
    \subsection{Survival Probability of single-particle states}
   Before comparing different paths of winding number transition, we introduce the concept of survival probability for the initial state. With this in mind, consider a single-particle system defined by Hamiltonian $\hat{H}$, which is a function of a parameter $\chi_{1} $ at $t < 0$. The single-particle states at this time are given by the Schr\"{o}dinger's equation, 
 \begin{eqnarray}
 H(\chi_{1})\ket{\psi_{i}(\chi_{1})}=E_{i}(\chi_{1})\ket{\psi_{i}(\chi_{1})}.
 \end{eqnarray}   
At $t=0$  a quench is performed, i.e., there is an instant change of parameter $\chi_{1} \rightarrow \chi_{2} $. The eigenstates $\ket{\psi_{f} }$ corresponding to the new Hamiltonian with the parameter $\chi_{2} $ are defined by  Schr\"{o}dinger's equation, 
\begin{eqnarray}
\hat{H}(\chi_{2})\ket{\psi_{f}(\chi_{2})}=E_{f}(\chi_{2})\ket{\psi_{f}(\chi_{2})}.
\end{eqnarray} 
The time evolution of a single particle state, $\ket{\psi_{i}(\chi_{1})}$, post-quench is given by
\begin{equation}\label{eq:11}	
	\ket{\psi_{i}(t)}=\sum_{f=1}^{N}e^{-iE_{f}(\chi_{2})t}\ket{\psi_{f}(\chi_{2})}\bra{{\psi_{f}(\chi_{2})}}\ket{\psi_{i}(\chi_{1})},
\end{equation}
where the sum is over the eigenstates associated with  $\hat{H}(\chi_{2})$. The likelihood of returning to our initial state at any time $t > 0$, also known as the survival probability  of the initial state and is given by: 	
\begin{eqnarray}
P^{i} (t)=\left|\bra{\psi_{i}(\chi_{1})}\ket{\psi_{i}(t)} \right|^{2}.
\end{eqnarray}
We will focus on this quantity in the first part of Section III to characterize the properties of a quench. More specifically, we will start with some well-defined edge state, characterized by the topology of the pre-quench system, and evaluate $P^i(t)$ post-quench.

    \section{Result and Discussions}
    
    \subsection{The dependence of survival probability on path}
\begin{figure}
  	\centering
  	\includegraphics[width=0.9\linewidth]{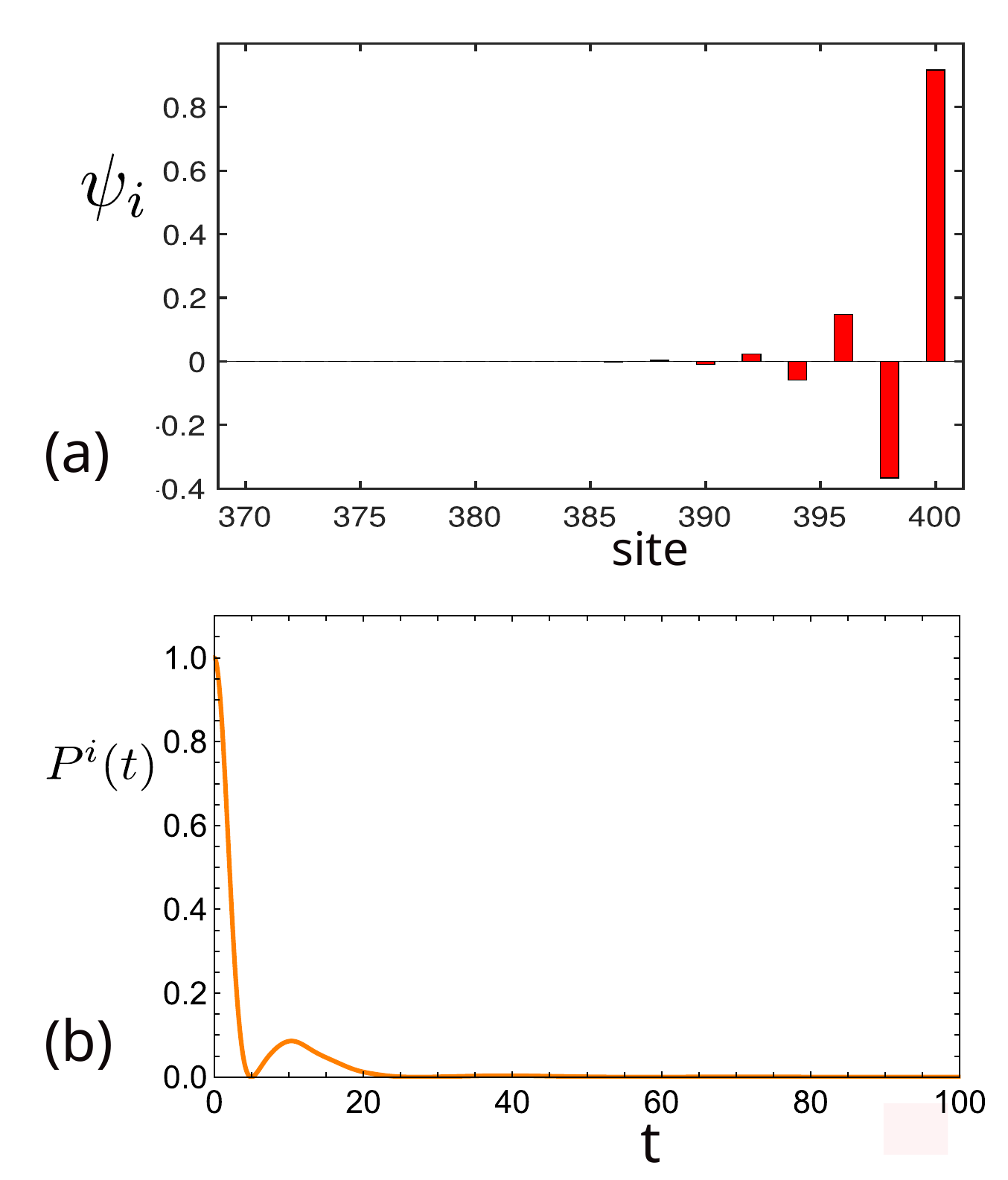}
  	\caption{(a) Edge state wave-function in a $N=400$ site system with $ v_i= 0.2 $ and $ w_i= 0.5 $. (b) Survival probability, $P^i(t)$, for $ v_i = 0.2 $, $ w_i = 0.5 $, $ v_f = 0.6 $ and $ w_f = 0.5 $ and the initial state is shown in (a).}
  	\label{fig:figure4}
	\vspace{-0.5cm}
  \end{figure}

    We know that edge states are topologically protected and the number of edge states is determined by the winding number of the system.  As we have discussed in previous sections, our system can host multiple topological phases, and there can be several paths possible for transition between the same set of winding numbers, see Fig.~\ref{fig:figure2}. Below we study the survival probabilities of edge states for quenches corresponding to different paths. As an example, we initially consider the standard SSH model with $\nu=\mu=\gamma=0$, and our initial system defined by $v_i<w_i$, specifically $ v_i = 0.2 $ and $ w_i = 0.5 $, with $C_i=1$. We then quench the system to $v_f>w_f$, specifically $v_f = 0.6 $ and $w_f = 0.5 $, with $C_f=0$. For such a scenario, we consider a system with $N=400$ sites, and the initial eigenstate is confined to the edge, see Fig.~\ref{fig:figure4}(a), with dimensionless eigenenergy $0$. After quenching, we see a dramatic drop off in the survival probability of the initial state \cite{sacramento2016edge, rajak2014survival}, see Fig.~\ref{fig:figure4}(b). The rationale for this rapid decay in survival probability is clear, i.e. we have quenched from a system, $C_i=1$ which supports an edge state to a system, $C_f=0$, which does not. 
  
\begin{figure}
  	\centering
  	\includegraphics[width=0.9\linewidth]{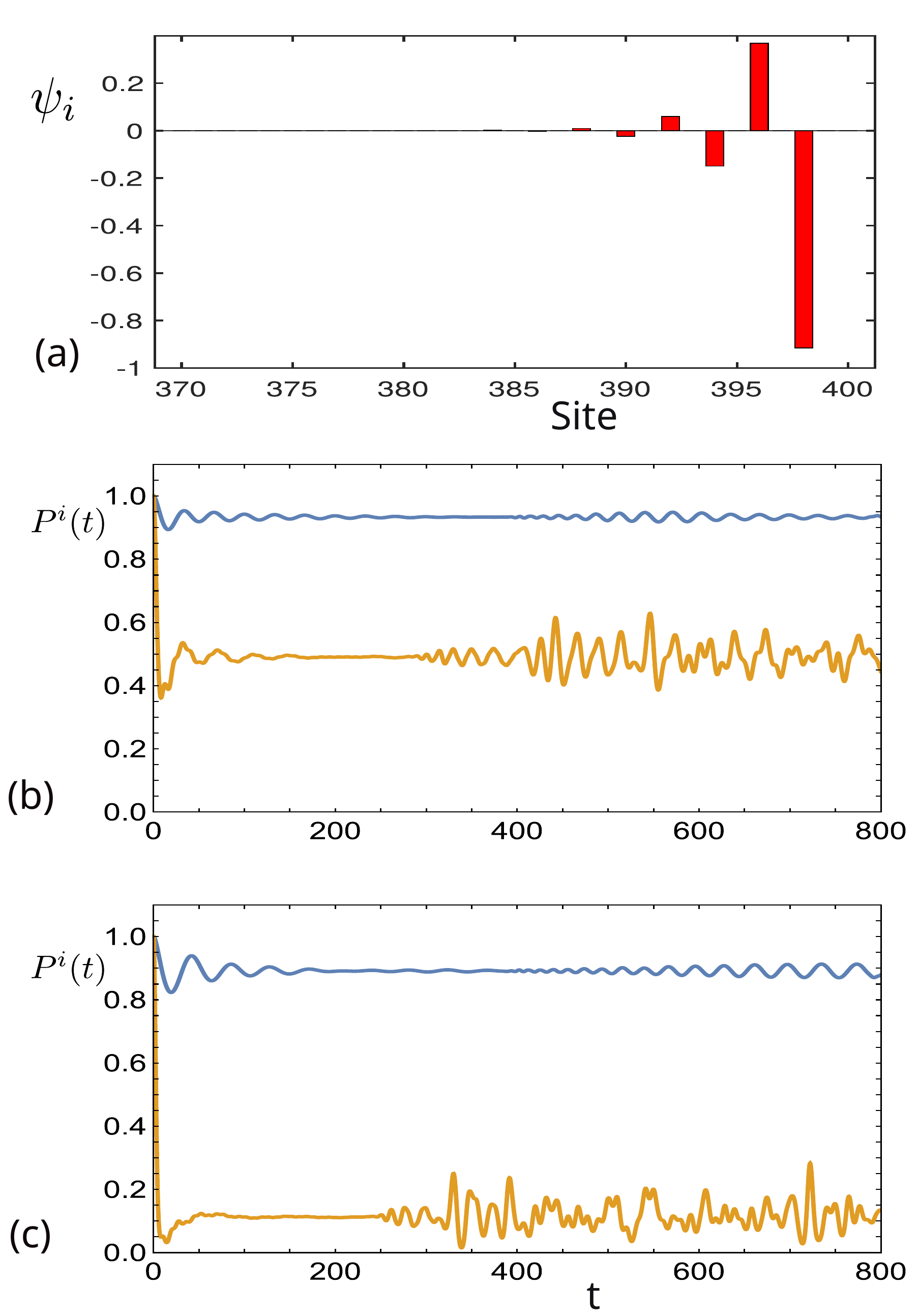}
  	\caption{ (a) Edge state wave-function in a $N=400$ site system with $v_i=0$, $w_i=0.17$, $\nu_i=0.43$, $\mu_i=0.17$ and $\gamma_i=0.37$, with $C_i=2$. (b) Overlap probability, $P^i(t)$, for $v_f=0.2$ (orange line) and $\mu_f=0.56$ (blue line). (c) Overlap probability, $P^i(t)$, for $v_f=0.3$ (orange line) and $\mu_f=0.6$ (blue line). In (b) and (c), the initial state considered is the one shown in (a).}
  	\label{fig:figure5}
	\vspace{-0.5cm}
  \end{figure}

In the following, we ask: if the post-quench system supports edge states, $C_f>0$, how will the survival probability of the initial edge state be affected, and how will the quench path impact it? As in Fig.~\ref{fig:figure4}, we consider a system with $N=400$ sites. The initial eigenstate we consider is shown in Fig.~\ref{fig:figure5}(a), with $v_i=0$, $w_i=0.17$, $\nu_i=0.43$, $\mu_i=0.17$ and $\gamma_i=0.37$, where $C_i=2$. In Fig.~\ref{fig:figure5}(b), the survival probability for this initial state is plotted along the two quench paths shown in Fig.~\ref{fig:figure2}, specifically, $v_f=0.2$ (orange line) and $\mu_f=0.56$ (blue line), noting that in each case $ C_f = 4 $.   
      
From Fig.~\ref{fig:figure5}(b), it appears that quenching into a state which supports edge states enhances the survival of the pre-quench edge state. However, we note that for the quench to $v_f=0.2$, the long-term behavior of the overlap with the initial state is just above $0.4$. This contrasts with the quench to $\mu_f=0.56$, where the long-term survival probability is just above $0.8$. Examining this more closely, we consider quenches $v_f=0.3$ (orange line) and $\mu_f=0.6$ (blue line), see Fig.~\ref{fig:figure5}(c). The quenches here correspond to those closer to the boundary of $C_f=4$. As can be seen in Fig.~\ref{fig:figure3}, if $v_f$ is increased much further, $C_f$ will eventually become $0$, whereas if $\mu_f$ is increased much further, $C_f$ will become $3$. As can be seen across Figs.~\ref{fig:figure5}(b) and (c), the long-term survival probability for the quench in $v$ is significantly less than for the quench in $\mu$. 
    
\begin{figure}
	\centering
	\includegraphics[width=0.9\linewidth]{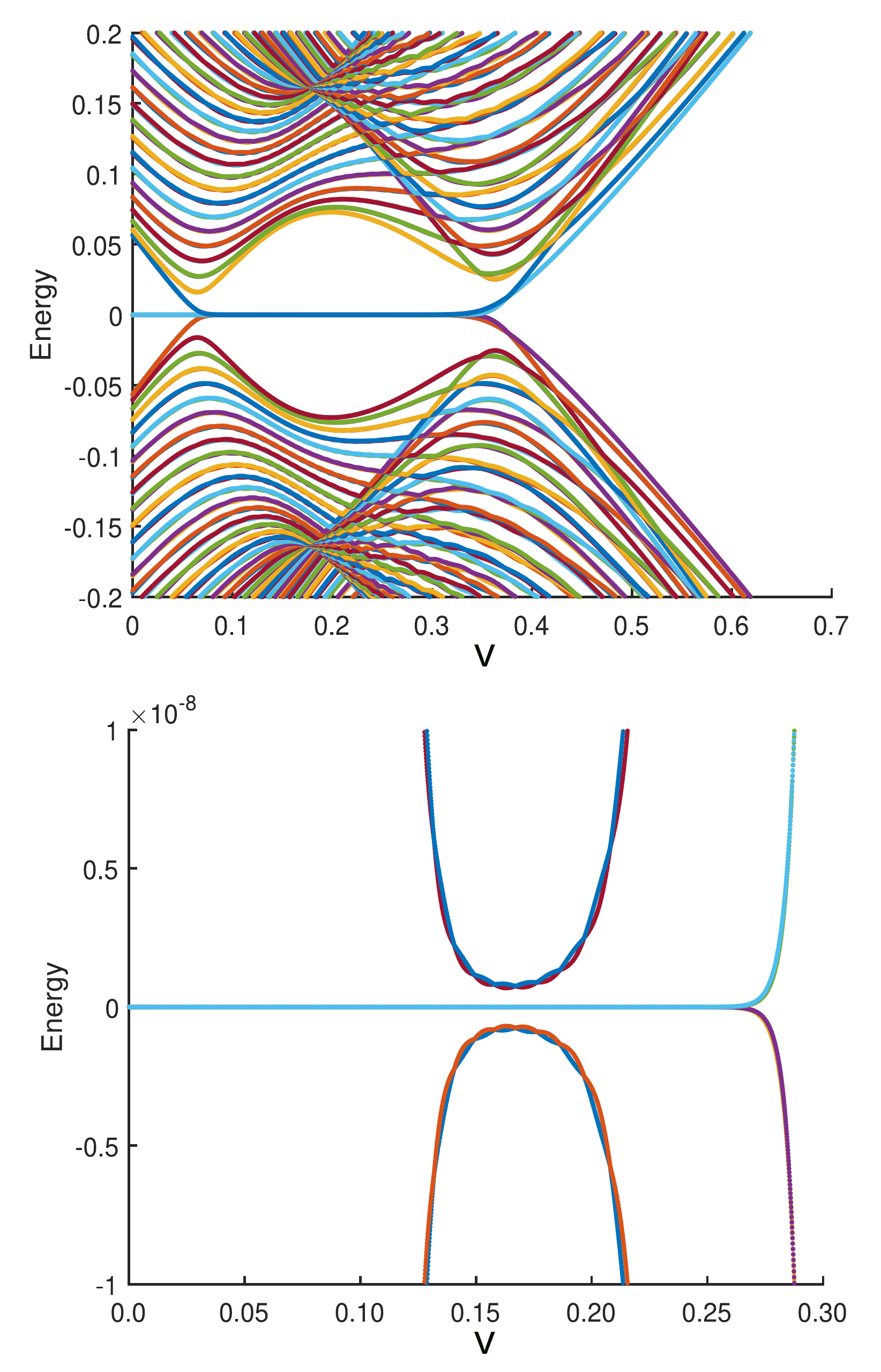}
	\caption{(a) The eigen-energies of a $N=400$ sites system for $w=0.17$, $\nu=0.43$, $\mu=0.17$ and $\gamma=0.37$ as a function of $v$. (b) A zoomed-in plot of (a) focusing on energies close to zero.}
	\label{fig:figure6}
	\vspace{-0.5cm}
\end{figure}

\begin{figure}
	\centering
	\includegraphics[width=0.9\linewidth]{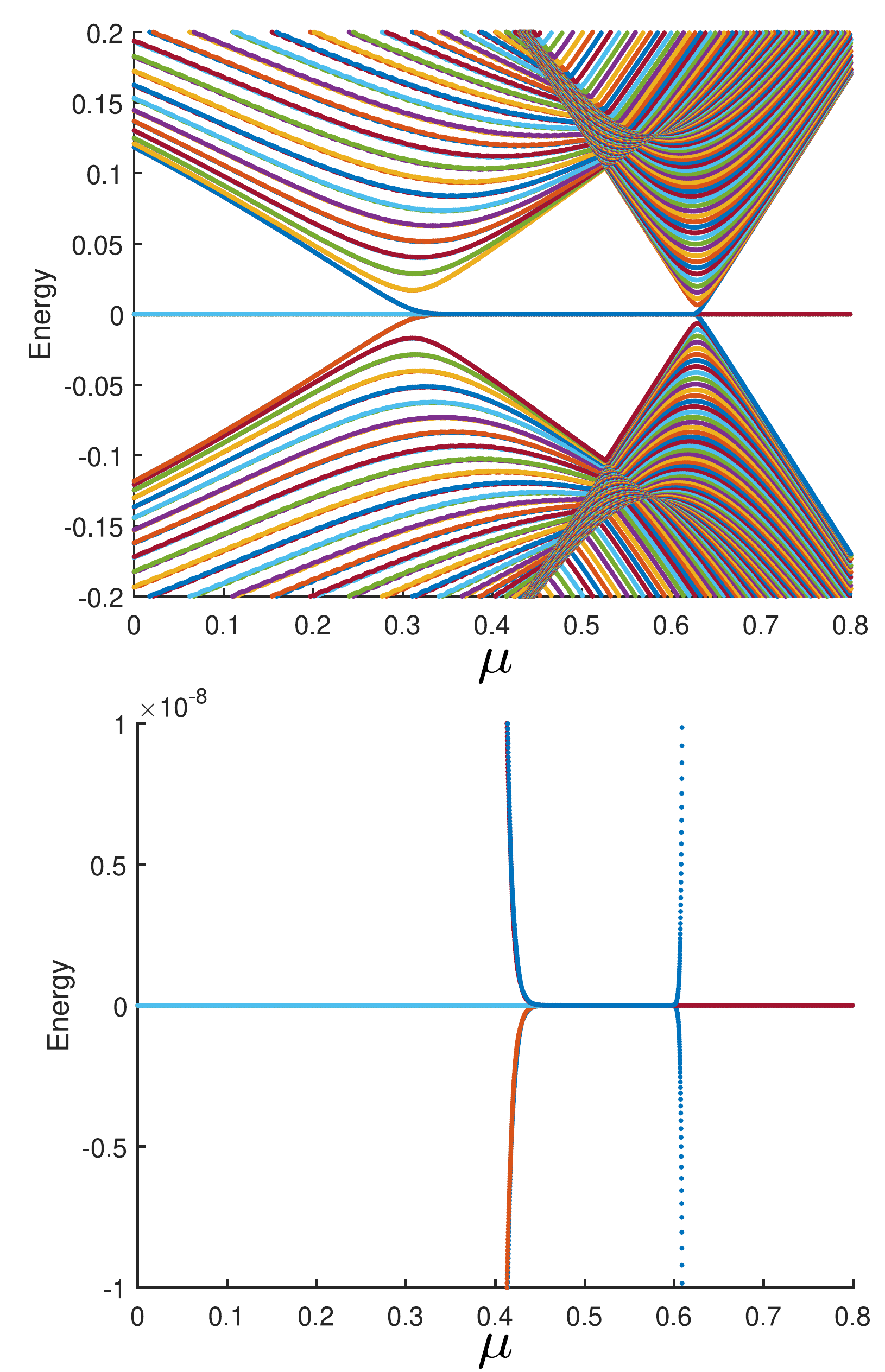}
	\caption{(a) The eigen-energies of a $N=400$ sites system for$v=0$, $w=0.17$, $\nu=0.43$ and $\gamma=0.37$ as a function of $\mu$. (b) A zoomed-in plot of (a) focusing on energies close to zero.}
	\label{fig:figure7}
	\vspace{-0.5cm}
\end{figure}

Other quenches show similar results, i.e., the survival probability of edge states is not only determined by the topology of the system they are quenched into but also by the direction in which the quench is heading. In the two examples, we have to consider that the paths can be labeled as $2-4-0$ (quenching $v$) and $2-4-3$ (quenching $\mu$). Writing the path labels in a more general sense as $C_i-C_f-C_h$, we can characterize the properties of the quench through examination of the energy band diagrams corresponding to the paths (Figs~\ref{fig:figure6} and \ref{fig:figure7}).

Figure \ref{fig:figure6} shows the eigenenergies for path $2-4-0$ as $v$ is increased. As can be seen, in Fig.~\ref{fig:figure6}(b), the {\it zero energy} states depart from their zero energy character just after the first topological phase transition~($ 2-4 $) as the parameter $v$ increases. This departure results in a decrease in the survival probability for the quench $2-4-0$, even though $ C_f = 4 $. However, the edge states maintain their character throughout the path $P2$ ($2-4-3$), see Fig.~\ref{fig:figure7}. This results in the survival probability of the edge state remaining relatively high after the quench.  

From these results, it is clear that the quench dynamics of some edge states depends on the path chosen for the topological phase transition, even if the final topology of the system, $C_f$, is the same. For the example considered, we see that the quench can be characterized as either $ 2-4-0 $ or $ 2 -4 - 3 $, and the latter path has a higher survival probability. From Figs.~\ref{fig:figure6} and ~\ref{fig:figure7}, we see that the details of the eigenenergy spectrum can provide insight into the survival probability. More generally, we have considered other quench paths and found that, for paths described by $C_i-C_f-C_h$, if two quenches are considered starting at the same initial state and having the same $C_f$ the survival probability will be larger if $C_h \ge C_i$ as compared to the case where $C_h < C_i$.

     In Fig. \ref{fig:figure5}, one can observe a pattern of the survival probability, which is related to interference. Specifically, in all cases, the survival probability, on short timescales, exhibits a dip. This is followed by a regime where the  survival probability is constant, and then comes a regime of ripples. We find that the time scale on which the ripples occur scales with system size. This can be understood by considering the transport of the probability density, with the ripples arising as a result of interference between probability density waves present at the edge and the reflected wave from the other edge. The timescale for the onset of this interference increases linearly with system size. The time of the arrival of the ripples in the survival probability is different for the two paths, and also, it is slightly different for \ref{fig:figure5} (b) and (c) corresponding to the paths. The underlying physics is related to the non-equilibrium transport of the edge states corresponding to the paths, which will be discussed in detail below. 
%   \begin{figure}
%   	\centering
%   	\includegraphics[width=0.7\linewidth]{"E:/My works/Figures/work 1/Figure 9"}
%   	\caption[Quench Dynamics vs. system size]{showing the dependence of the s.p of an edge state with system size in a quench from the configuration set [0.0, 0.2, 0.15] to [0.1, 0.2, 0.15]. These are the s.p vs. time graphs and parts (a), (b), (c) and (d) corresponds to the number of lattice sites (Nm) 400,200,100 and 50 respectively.}
%   	\label{fig:figure-9}
%   \end{figure}
\\
\subsection{Variation of transport velocity with the type of path}
\begin{figure}
	\centering
	\includegraphics[width=0.9\linewidth]{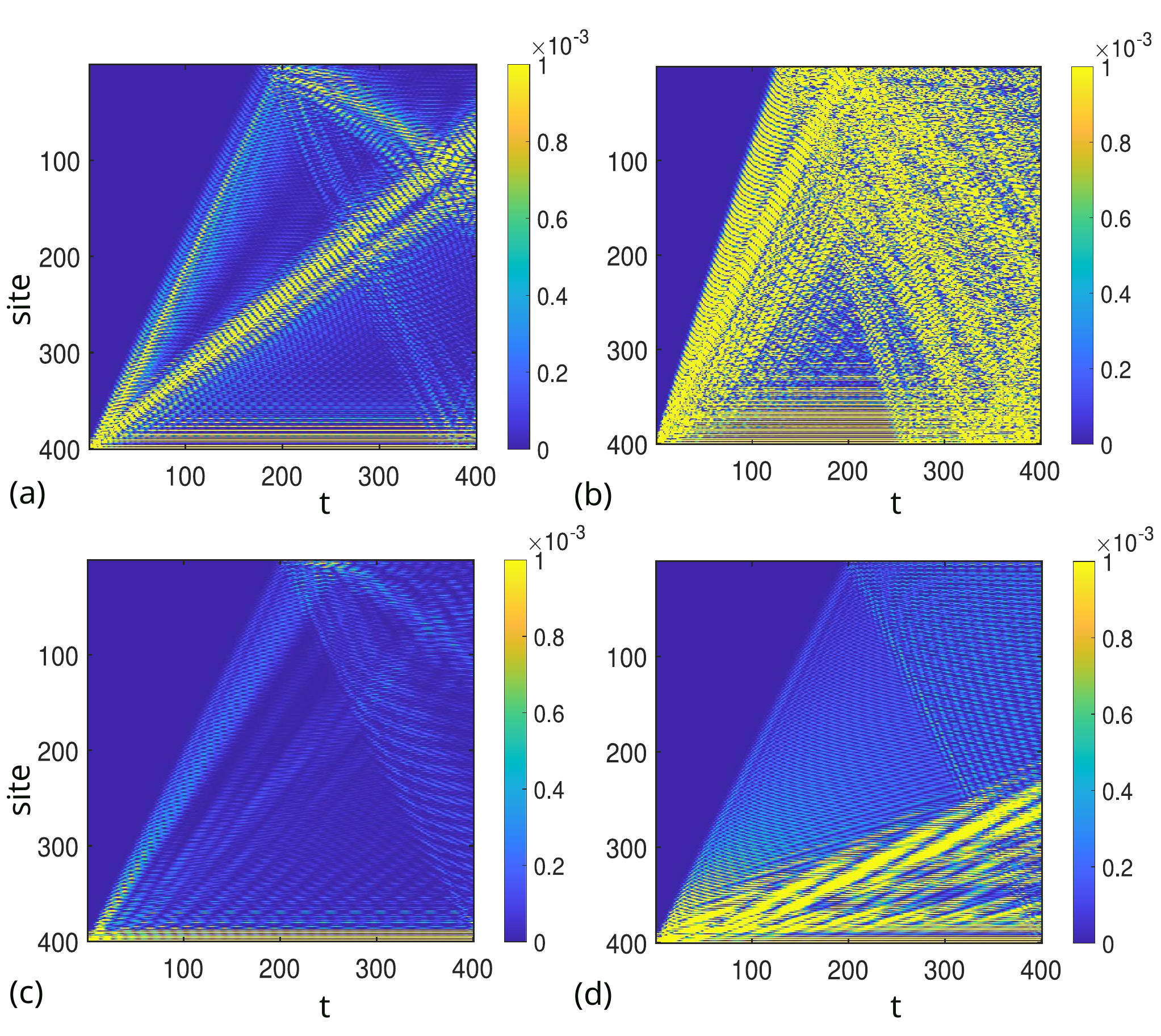}
	\caption{Four light-cone diagrams for quenches of $v$ from $v_i=0$ to (a) $v_f=0.1$ and (b) $v_f=0.3$ with $w$, $\nu$, $\mu$ and $\gamma$ fixed at $0.17$, $0.43$, $0.17$ and $0.37$ respectively and quenches of $\mu$ from $\mu_i=0.17$ to (c) $\mu_f=0.39$ and (b) $\mu_f=0.59$ with $v$ $w$, $\nu$ and $\gamma$ fixed at $0$, $0.17$, $0.43$ and $0.37$ respectively. Each light-cone diagram plots the probability density ($n_i(t)== \bra{\psi_{i}(t)}\ket{\psi_{i}(t)}$ ) on each site $i$ (y-axis) as a function of time (x-axis).}
	\label{fig:figure8}
	\vspace{-0.5cm}
\end{figure}

The discussion in the previous subsection suggests that the post-quench dynamics are associated with the transport of the edge states across the system. Below we visualize this transport using light-cone diagrams, i.e., a diagram that plots probability density ($n_i(t)= \bra{\psi_{i}(t)}\ket{\psi_{i}(t)} $) on each site with respect to time. Fig. \ref{fig:figure8} characterizes the transport of the edge state for quenches along $P1$ corresponding to the values $v_{f}=0.1$ and $v_{f}=0.3$ in (a) and (b), respectively, with $v_i=0$. Whereas, Figs. \ref{fig:figure8} (c) and (d) correspond to the transport of the edge state for quenches along $P2$ corresponding to the values $\mu_{f}=0.39$ and $\mu_{f}=0.59$ respectively, with $\mu_i=0.17$. Comparing Figs.  \ref{fig:figure8} (a) with (b),  and (c) with (d), it can be observed that there is a change in the velocity of the channels through which the maximum transport is taking place as the final quench configuration is moved along $P1$ and $P2$ respectively. This velocity variation is quantitatively different for the two paths of the quench. By comparing Figs.  \ref{fig:figure8} (a) with (b), we see that the velocity of the maximum transport increases for $P1$. More generally we find that the velocity of maximum transport channels increases for paths that characterize a decaying survival probability of edge states. Whereas for the path which does not characterize a decaying survival probability of edge state ($P2$), the velocity variation of the channel of maximum transport is from high to low as we move along the quench path, as can be seen when comparing Figs.  \ref{fig:figure8} (c) with (d). Close examination of Fig. \ref{fig:figure8}(a) reveals that there are two dominant channels through which the transport takes place. One of these channels corresponds to a relatively high velocity (lower intensity) as compared to the other, which corresponds to a lower velocity (higher intensity). As the final quench configuration is moved along the path $P1$, the velocity corresponding to both of the channels increases. Also, during the journey along path $P1$, the velocities tend to be closer in magnitude, and the intensity of the faster velocity channel increases slightly. On the other hand, Fig.~\ref{fig:figure8} (c) demonstrates that there is very low transport in total. As the final quench configuration is moved along path $P2$, there is an increased transport along a much lower velocity channel than the previous configuration.
 
\begin{figure}
	\centering
	\includegraphics[width=0.9\linewidth]{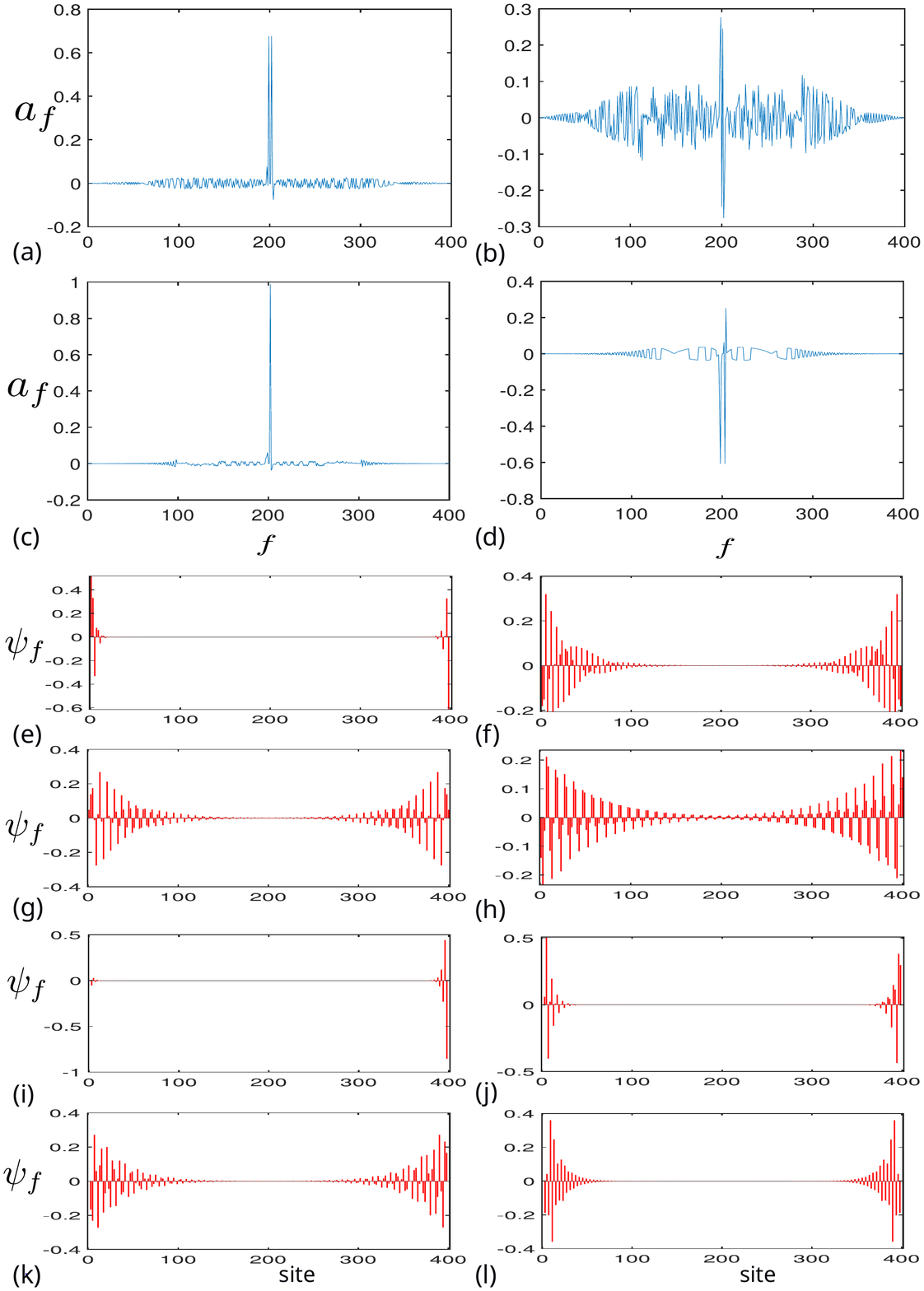}
	\caption{In this Fig., (a), (b), (c), and (d) are the plots of the coefficient $  a_f=\langle \psi_f(\chi_2)|\psi_i(\chi_1)\rangle  $ against the indices f corresponding to $ v_{f} = 0.1 $, $ v_{f}=0.3 $ of P1 and $ \mu_{f}=0.39 $, $ \mu_{f}=0.59 $ of P2 respectively. (e), (g), and (f),(h) correspond to edge states of $ v_{f} = 0.1 $ and $ v_{f} = 0.3 $ along P1, respectively, with (e),(f) having the lowest order of energies and (g),(h) having relatively higher orders of energies in their respective Hamiltonians. (i), (k), and (j),(l) correspond to edge states of $ \mu_{f} = 0.39 $ and $ \mu_{f} = 0.59 $ along P2, respectively, with (i), (j) having the lowest order of energies and (k), (l) having a relatively higher order of energies in their respective Hamiltonians.
		 The central peaks from 197 to 204 in the coefficient plots correspond to the high overlap with some of the edge states of the final Hamiltonian. The delocalization in the edge states along P1 is clear if we compare (e), (g) with (f), (h). Similarly, the localization in the transport contributing edge state along P2 is apparent if we compare (k) with (l).}
	\label{fig:figure9}
	\vspace{-0.5cm}
\end{figure}

In order to explain the velocity variation phenomenon, we explore the correspondence between the components of Eq.~(\ref{eq:11}) and the light cones depicted in Fig.\ref{fig:figure8}. We examine, in Fig.~\ref{fig:figure9}, the change in the coefficients, $a_f=\langle \psi_f(\chi_2)|\psi_i(\chi_1)\rangle$, of the basis vectors in the expression for the evolving state in Eq.~(\ref{eq:11}),  and the spatial structure of the edge states as we move along the path i.e., the localization and delocalization of the edge states. The coefficient $ a_f $ is dependent on the spatial structure of the eigenstates of the final Hamiltonian. Since the summation in Eq.~(\ref{eq:11}) is over all the eigenstates of the final Hamiltonian, both the edge states and the bulk states contribute to the transport. The coefficient of terms in the summation dictates its weight and hence corresponds to the intensity of the channel. The energies of states, along with the time, come as a phase factor in the expression and hence contributes to the velocity of transport. Figures \ref{fig:figure9}(a-d)  correspond to the coefficients for $v_f=0.1 $, $ v_f=0.3 $ of path $ P1 $ and $ \mu_f = 0.39 $, $ \mu_f = 0.59 $ of path $ P2 $ respectively. The coefficients from 197 to 204 in Figs. \ref{fig:figure9} (a-d) correspond to the edge states and the rest to the bulk states. The peaks in the central region of these figures come from the strong overlap of the initial edge state with the edges states in the post-quench Hamiltonian. In Fig.\ref{fig:figure9} (a-d), the highest peaks correspond to the edge states with the lowest order of energy (as shown in Figs.\ref{fig:figure9}(e),(f),(i), and (j) for $ v_f=0.1 $ and $ v_f=0.3$ ($P1$), $ \mu_f = 0.39 $ and $ \mu_f = 0.59 $ ($P2$) respectively and these contribute to the static horizontal lines at the bottom of the light-cone diagrams. It may be observed that the spread of the horizontal lines toward the center of the wire is more in Fig.\ref{fig:figure8}(b) than in Fig.\ref{fig:figure8}(a), and the reason for this may be attributed to the fact that there is a delocalization in the edge states as we move from $ v=0.1 $ to $ v=0.3 $. The delocalization is clearly visible if we compare Fig.9(e),(g) with Fig.9(f),(h) which corresponds to $ v=0.1 $ and $ v=0.3 $ of $ P1 $ respectively. For all of the cases, the next highest contribution comes from the edge states whose energies are close to zero but are orders of magnitude higher than that of the stationary edge states and these correspond to the slow velocity channel in the light cones of Figs.\ref{fig:figure8}. The energies of these states increase as we move along the paths, and this results in an increase in the velocity of the slower velocity channel. This result is more prominent in Figs.\ref{fig:figure8}(a) and (b) as the corresponding coefficients in this path $ P1 $ are higher than in $ P2 $. As we move along the path $P2$ (Figs.\ref{fig:figure8}(c) and (d)), it might appear that the velocity of the channel of most transport decreases, as shown in Figs. \ref{fig:figure8}(c) and (d), however, it appears so because the coefficients of the lower velocity channel were initially very low and it increased as we moved along the path, resulting in significant transport through the low-velocity channel, making the lower velocity channel most intense. It is important to note that the coefficient corresponding to the lower velocity channel increases along path P2 due to the localization of the edge states. Due to localization, the amplitude near the edges increases, increasing the coefficient. 
 The coefficients corresponding to the bulk-states in Fig.\ref{fig:figure9}(b) are much higher when compared to the coefficients in Fig. \ref{fig:figure9}(a). This is the reason that the intensity of the higher velocity channel in the former is much higher than that of the latter. The velocity of the higher velocity channel also increases as we move along a path, and it is because of the increase in the bulk state energies. As we move along the path $ P2 $, the bulk state energies increase as well, but there is localization in the edge states, leading to less contribution to transport, and so the velocity and magnitude of the higher velocity channel do not increase significantly.
 
   \section{Summary and Conclusions}
By analyzing the topological phases of the extended SSH model using winding number diagrams we have shown that there can be multiple paths of winding number transition possible for a quench between two topological phases. By studying the path dependence of quench dynamics in terms of fermionic edge-state survival probability we found that it is possible to classify the survival probability in terms of three general parameters: (i) the pre-quench winding number ($C_i$); (ii) the post-quench winding number ($C_f$) and (iii) the winding number of where the quench is heading ($C_h$). In general, we find that for a system characterized by $Ci-C_f-C_h$ if two quenches are considered with the same $C_i$ and $C_f$, the survival probability for some of the fermionic edge-state with be greater for the quench with $C_h \ge C_i$ as compared to the case where $C_h<C_i$.  This classification of the robustness of edge states when quenching between two topological regimes has only been tested within this model and further work is required to test this concept more broadly.  We then studied post-quench dynamics through the use of light cone diagrams associated with probability density. Focusing on the example paths we have considered, we found that the velocities of the channels of most transport vary along the path of the winding number transition. This variation depends on the path we choose, i.e., for the path characterizing a decaying (stable) survival probability, the velocity corresponding to the most transport increases (decreases) along the path. The velocity variation phenomenon is explained by analyzing the coefficient plots, the energy spectrum of the final Hamiltonian, and the spatial structure of the edge states of the final Hamiltonian.

	\section*{ACKNOWLEDGMENTS}
We want to express our great appreciation to all our group members at the University of Melbourne and IIT Kharagpur for their valuable and constructive suggestions during the planning and development of this research work. Their willingness to participate eagerly in the discussion sessions is greatly appreciated.
This project was supported by an Institute fellowship from IIT Kharagpur, the government of India, and N.D. Goldsworthy scholarship from the University of Melbourne, Australia. We are deeply thankful to them for supporting this project.

\bibliographystyle{unsrtnat}
\bibliography{References}
\end{document}